\documentclass[preprint,prd,superscriptaddress]{revtex4}

\usepackage{graphicx}% Include figure files
\usepackage{dcolumn}% Align table columns on decimal point
\usepackage{bm}% bold math
\newcommand{\p}{\partial}
\newcommand{\pslash}{p\kern-1ex /}
\newcommand{\lslash}{l\kern-1ex /}
\newcommand{\kslash}{k\kern-1ex /}
\newcommand{\dslash}{\p\kern-1.2ex /}
\newcommand{\Dslash}{{\cal D}\kern-1.5ex /}
\newcommand{\Aslash}{A\kern-1.2ex /}

\begin{document}

\preprint{RIKEN-TH-91}
\preprint{UTHEP-538}
\preprint{KEK-CP-191}
\preprint{YITP-07-4}
\preprint{NTUTH-07-505A}

\title{
  Two-flavor lattice QCD simulation in the $\epsilon$-regime
  with exact chiral symmetry
}

\newcommand{\RIKEN}{
  Theoretical Physics Laboratory, RIKEN,
  Wako 351-0198, Japan
}

\newcommand{\Taiwan}{
  Physics Department and Center for Theoretical Sciences, 
                National Taiwan University, Taipei, 10617, Taiwan  
}

\newcommand{\Tsukuba}{
  Graduate School of Pure and Applied Sciences, University of Tsukuba,
  Tsukuba 305-8571, Japan
}
\newcommand{\BNL}{
  Riken BNL Research Center, Brookhaven National Laboratory, Upton,
  NY11973, USA
}
\newcommand{\KEK}{
  High Energy Accelerator Research Organization (KEK),
  Tsukuba 305-0801, Japan
}
\newcommand{\GUAS}{
  School of High Energy Accelerator Science,
  The Graduate University for Advanced Studies (Sokendai),
  Tsukuba 305-0801, Japan
}

\newcommand{\YITP}{
  Yukawa Institute for Theoretical Physics, 
  Kyoto University, Kyoto 606-8502, Japan
}

\newcommand{\HUDP}{
  Department of Physics, Hiroshima University,
  Higashi-Hiroshima 739-8526, Japan
}

\author{H.~Fukaya}
\affiliation{\RIKEN}

\author{S.~Aoki}
\affiliation{\Tsukuba}
\affiliation{\BNL}

\author{T.W.~Chiu}
\affiliation{\Taiwan}

\author{S.~Hashimoto}
\affiliation{\KEK}
\affiliation{\GUAS}

\author{T.~Kaneko}
\affiliation{\KEK}
\affiliation{\GUAS}

\author{H.~Matsufuru}
\affiliation{\KEK}

\author{J.~Noaki}
\affiliation{\KEK}

\author{K.~Ogawa}
\affiliation{\Taiwan}

\author{M.~Okamoto}
\affiliation{\KEK}

\author{T.~Onogi}
\affiliation{\YITP}

\author{N.~Yamada}
\affiliation{\KEK}
\affiliation{\GUAS}

% \author{Hidenori~Fukaya}
% \affiliation{\RIKEN}

% \author{Sinya Aoki}
% \affiliation{\Tsukuba}
% \affiliation{\BNL}

% \author{Ting-Wai Chiu}
% \affiliation{\Taiwan}

% \author{Shoji~Hashimoto}
% \affiliation{\KEK}
% \affiliation{\GUAS}

% \author{Takashi~Kaneko}
% \affiliation{\KEK}
% \affiliation{\GUAS}

% \author{Hideo~Matsufuru}
% \affiliation{\KEK}

% \author{Kenji Ogawa}
% \affiliation{\Taiwan}

% \author{Masataka Okamoto}
% \affiliation{\KEK}

% \author{Tetsuya~Onogi}
% \affiliation{\YITP}

% \author{Norikazu~Yamada}
% \affiliation{\KEK}
% \affiliation{\GUAS}

\collaboration{JLQCD collaboration}
\noaffiliation


\begin{abstract}
  We perform lattice simulations of two-flavor QCD 
  using Neuberger's overlap fermion, with which the exact
  chiral symmetry is realized at finite lattice spacings.
  The $\epsilon$-regime is reached by decreasing the light quark
  mass down to 3~MeV on a $16^3\times 32$ lattice with
  a lattice spacing $\sim$ 0.11~fm.
  We find a good agreement of the low-lying Dirac eigenvalue
  spectrum with the analytical predictions of the chiral
  random matrix theory, which reduces to the chiral
  perturbation theory in the $\epsilon$-regime.
  The chiral condensate is extracted as
  $\Sigma^{\overline{\mathrm{MS}}}(2\mathrm{~GeV})$ = 
  $(251 \pm 7 \pm 11 \mathrm{~MeV})^3$,
  where the errors are statistical and an estimate of the higher
  order effects in the $\epsilon$-expansion.
\end{abstract}

\maketitle
In Quantum Chromodynamics (QCD), it is widely believed
that chiral symmetry is spontaneously broken, making
pions nearly massless while giving masses of order
$\Lambda_{\mathrm{QCD}}$, the QCD scale, to the other
hadrons. 
In fact, the Chiral Perturbation Theory (ChPT), an effective
theory based on the spontaneously broken chiral symmetry,
describes low energy interactions of pions very
accurately.  
Nevertheless, theorists have not been successful in
analytically solving QCD and deriving the chiral symmetry
breaking, due to its highly non-perturbative dynamics. 

The most promising approach to establishing the link
between QCD and ChPT is to utilize the numerical simulation
of lattice QCD, with which the every prediction of ChPT can
be tested in principle.
For instance, the presence of the so-called chiral
logarithms, the effect of pion cloud, should be reproduced.
Such a numerical test is, however, not an easy task, because
of rapidly increasing computational cost in the small quark
mass region where ChPT is reliably applied.
Another serious problem is the explicit violation of the
chiral symmetry at finite lattice spacings in the
conventional fermion formulations, with which the conclusive
test of ChPT requires well-controlled and thus
computationally demanding continuum extrapolation.

In this work we improve this situation in two ways.
First, we employ Neuberger's overlap fermion 
%\cite{Narayanan:1994gw,Neuberger:1997fp,Neuberger:1998wv} 
\cite{Neuberger:1997fp,Neuberger:1998wv} 
for dynamical quarks.
It preserves exact chiral symmetry at finite lattice
spacings, and hence ChPT can be applied before
taking the continuum limit.
Although the numerical cost of the overlap fermion
is almost 100 times higher than that of the other fermions,
new computational facilities at KEK enable us to carry out
such a work. 

Second, we study the correspondence between QCD and ChPT in
the so-called $\epsilon$-regime
\cite{Gasser:1987ah,Hansen:1990un,Hansen:1990yg}, which is
characterized by the small pion mass $m_\pi$ satisfying 
$m_\pi L \lesssim 1$ with $L$ the box size.
In this regime ChPT is safely applied as an expansion in 
terms of $\epsilon^2\sim m_\pi/\Lambda_{\mathrm{QCD}}$,
provided that the condition 
$1/(\Lambda_{\mathrm{QCD}}L)^2\ll 1$, the usual condition
that the box size is larger than the inverse QCD scale, is
satisfied. 
We set the sea quark mass to $\sim$ 3~MeV, for which 
$m_\pi L\simeq$ 1.0.
With the space-time volume 
$L^3\times T\simeq (1.8~\mathrm{fm})^3\times(3.5~\mathrm{fm})$, 
the numerical cost is still not prohibitive even with such a
small sea quark mass, since the finite volume provides a
natural lower bound on the lowest eigenvalue of the Dirac
operator. 

In the $\epsilon$-regime, 
zero-momentum modes of the pion field dominate the dynamics
and the kinetic term gives only subleading contributions.
The Lagrangian of ChPT reduces to 
$\mathcal{L}^{(0)}=m\Sigma\,\mathrm{Re} \mathrm{Tr}[U]$
with $m$ the (degenerate) quark mass 
and $U$ ($\in$ SU($N_f$)) the pion field ($N_f=2$ in this
work). 
The system is fully characterized by a parameter 
$m\Sigma V$, where $\Sigma$ is the chiral condensate and $V$
is the space-time volume $L^3\times T$. 
Dependence on the topological charge $Q$ of the gauge field
also becomes significant. 

At the leading order of the $\epsilon$-expansion, 
the chiral Random Matrix Theory (ChRMT)
provides an equivalent description of ChPT
\cite{Shuryak:1992pi,Verbaarschot:1993pm,Damgaard:2000ah}. 
Furthermore, ChRMT can predict the distributions of the
individual eigenvalues of the Dirac operator, which may 
be directly compared with the lattice data.
In the quenched approximation, a good agreement between ChRMT
and lattice calculation has been observed using the
overlap-Dirac operator 
\cite{Edwards:1999ra,Bietenholz:2003mi,Giusti:2003gf}, and
$\Sigma$ has been determined by matching the eigenvalues
\cite{Wennekers:2005wa}. 
The present work is an extension of these works to two-flavor
QCD.
A preliminary report of this work has been presented in
\cite{Fukaya:2006xp}, and an overview of our dynamical
overlap fermion project is found in \cite{Kaneko:2006pa}.
Similar studies have been done recently 
\cite{DeGrand:2006nv,Lang:2006ab}, but the $\epsilon$-regime
was not reached because of larger sea quark masses.

We have performed numerical simulations on a $16^3\times 32$ 
lattice at a lattice spacing $a\sim$ 0.11~fm as determined
from the scale $r_0$($=$0.49fm) of the heavy quark potential. 
We employ the overlap fermion
\cite{Neuberger:1997fp,Neuberger:1998wv}, whose Dirac 
operator is 
\begin{equation}
  \label{eq:overlap-Dirac}
  D(m) = 
  \left(m_0+\frac{m}{2}\right)+
  \left(m_0-\frac{m}{2}\right)
  \gamma_5 \mathrm{sgn}[H_W(-m_0)]
\end{equation}
for the quark mass $m$.
Here, $H_W(-m_0)$ denotes the standard hermitian Wilson-Dirac
operator $H_W(-m_0)\equiv\gamma_5D_W(-m_0)$ with a large
negative mass term (we choose $m_0$ = 1.6 throughout this work).
For the gauge part, the Iwasaki action is used at 
$\beta$ = 2.35 together with unphysical Wilson fermions and
associated twisted-mass ghosts \cite{Fukaya:2006vs}, 
which preserves the global topological charge during
molecular-dynamics evolutions of the gauge field. 
This is desirable for the $\epsilon$-regime simulations 
since we can effectively accumulate statistics at a given
topological charge. 
In this work our simulation is confined in
a topological sector $Q=0$.

For the simulation with the dynamical overlap fermions \cite{Fodor:2003bh},
we use the Hybrid Monte Carlo (HMC) algorithm.
The sign function in (\ref{eq:overlap-Dirac}) is
approximated by a rational function with Zolotarev's optimal
coefficients %\cite{vandenEshof:2002ms}
after projecting out low-lying eigenvalues of $|H_W(-m_0)|$. 
With 10 poles the sign function has a $10^{-(7-8)}$ precision.
Thanks to the extra Wilson fermions, the lowest
eigenvalue of $H_W(-m_0)$ never passes zero, and hence no
special care of the discontinuity of the fermion determinant is
needed. 
%\cite{Fodor:2003bh, Cundy:2004pz}

The simulation cost is substantially reduced by the mass
preconditioning of the HMC Hamiltonian
\cite{Hasenbusch:2001ne}.
The heavier overlap fermion mass for the preconditioner is
chosen to 0.4 except for two lightest sea quark masses where
the value is 0.2.
The relaxed conjugate gradient algorithm to invert the
overlap-Dirac operator \cite{Cundy:2004pz} also helps to
speed up the simulation by about a factor of 2.

\begin{figure}[tb]
  \centering
  \includegraphics[width=8cm,clip=true]{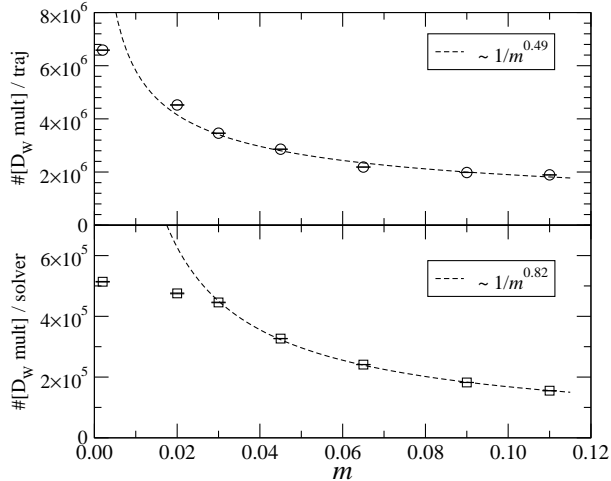}
  \caption{
    Number of the Wilson-Dirac operator multiplication per
    trajectory (upper panel) and per an overlap inversion
    (lower panel).
    The curves are fit to data above $m$ = 0.030 with the
    form $\propto 1/m^\alpha$.
  }
  \label{fig:cost}
\end{figure}

For the sea quark mass we take seven values: 0.110, 0.090,
0.065, 0.045, 0.030, 0.020, and 0.002.
The lightest sea quark ($m$ = 0.002) corresponds
to the $\epsilon$-regime. 
The simulation cost measured by the number of the
Wilson-Dirac operator multiplication is plotted in
Fig.~\ref{fig:cost}. 
The upper panel shows the cost per trajectory of length 0.5;
the lower panel presents the cost of inverting the overlap-Dirac
operator when we calculate the Hamiltonian at the end of
each trajectory. 
Increase of the numerical cost towards the chiral limit is
not as strong as expected: 
if we fit the data assuming the scaling law 
$\sim 1/m^\alpha$ above $m$ = 0.030, the power $\alpha$ is
0.82 for the inversion and 0.49 for the total cost.
In the $\epsilon$-regime the cost is even lower than the
expectation from the power law.
This is because the cost is governed by the lowest-lying
eigenvalue rather than the quark mass as explained below.

The number of trajectories is 1,100 for each sea quark mass
after discarding 500 for thermalization. 
For the $\epsilon$-regime run at $m$ = 0.002, we have
accumulated 4,600 trajectories. 
 In addition, we have generated quenched lattices at a
  similar lattice spacing with $Q=0$ and 2.
The computational cost at $m=0.002$ is about one hour per
  trajectory on a half rack (512 nodes) of the IBM BlueGene/L
(2.8~TFlops peak performance).

The lowest 50 eigenvalues of the overlap-Dirac
operator $D(0)$ are calculated at every 10 trajectories.
We use the implicitly restarted Lanczos algorithm for a
chirally projected operator $P_+\,D(0)\,P_+$, where
$P_+\!=\!(1+\gamma_5)/2$. 
From its eigenvalue ${\rm Re} \lambda^{ov}$
the pair of eigenvalues $\lambda^{ov}$ (and its complex conjugate) 
of $D(0)$ is extracted through the relation 
$|1-\lambda^{ov}/m_0|^2=1$, that forms a circle on a complex plane. 
%The eigenvalues of $D(0)$ form a circle on the complex plane while
%the continuum Dirac eigenvalues are pure imaginary.
For the comparison with ChRMT, the lattice eigenvalue
$\lambda^{ov}$ is projected onto the imaginary axis as
$\lambda\equiv\mathrm{Im}\lambda^{ov}/(1-\mathrm{Re}\lambda^{ov}/(2m_0))$.
Note that $\lambda$ is very close to
$\mathrm{Im}\lambda^{ov}$ (within 0.05\%) for the low-lying
modes we are interested in. 

\begin{figure}[tb]
  \centering
  \includegraphics[width=7cm,clip=true]{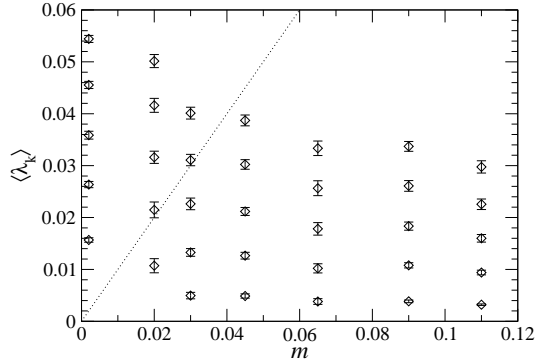}
  \caption{
    Lowest 5 eigenvalues $\lambda_k$ as a function of sea
    quark mass.
    Dashed line shows $\lambda=m$.
  }
  \label{fig:lowest_mdep}
\end{figure}

In Fig.~\ref{fig:lowest_mdep} we plot the ensemble averages
of the lowest 5 eigenvalues $\langle\lambda_k\rangle$ ($k$ =
1--5) as a function of the sea quark mass.
We observe that the low-lying spectrum is lifted
as the chiral limit is approached.
This is a direct consequence of the fermion determinant
$\sim \prod_k (|\lambda_k|^2+m^2)$, which repels the small
eigenvalues from zero when $m$ becomes smaller than the
lowest eigenvalue.
This is exactly the region where the numerical cost
saturates as it is controlled by $\lambda_1$ rather than $m$. 
We also find that the autocorrelation length of the lowest
eigenvalues is significantly longer in the $\epsilon$-regime.
%We therefore take the jackknife bin size of the statistical
%analysis to be 200 trajectories for the run with $m$ = 0.002
%while keeping it 100 trajectories for heavier sea quark runs.
We therefore use the jackknife method in the statistical
analysis with a bin size of 200 trajectories, with which
the statistical error saturates.

In the $\epsilon$-regime, ChRMT predicts the probability 
distribution $p_k(\zeta_k)$
of lowest-lying eigenvalues $\zeta_k$, 
and thus their ensemble averages 
$\langle\zeta_k\rangle = \int_0^\infty d\zeta_k \zeta_k p_k(\zeta_k)$.
The correspondence between ChRMT and ChPT implies the
relation 
$\langle\zeta_k\rangle = \langle\lambda_k\rangle\Sigma V$,
from which we can extract $\Sigma$, once
$\langle\lambda_k\rangle$'s are obtained.
Correction for finite sea quark mass $m$ can also 
be calculated by ChRMT.
At $m$ = 0.002, it is only 0.9\% for the lowest
eigenvalue, which is taken into account in the following
analysis. 

\begin{figure}[tb]
  \centering
  \includegraphics[width=8cm]{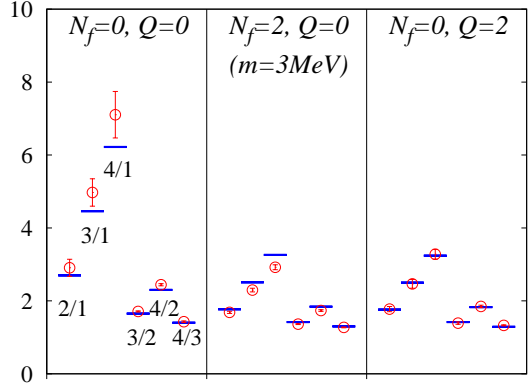}
  \caption{
    Ratio of the eigenvalues 
    $\langle\zeta_k\rangle/\langle\zeta_l\rangle$
    for combinations of $k$ and $l$ $\in$ 1--4
    (denoted in the plot as $k/l$).
    In addition to the two-flavor QCD data (middle),
    quenched data at $Q=0$  (left) and 2
    (right) are also shown.
    Lattice data (circles) are compared with
    the ChRMT predictions (bars).
  }
  \label{fig:eigenratio}
\end{figure}

We first compare the pattern of the eigenvalue spectrum
of the Dirac operator.
In the ratios 
$\langle\zeta_k\rangle/\langle\zeta_l\rangle$
of low-lying eigenvalues the factor $\Sigma V$ drops off and
the comparison is parameter free.
As Fig.~\ref{fig:eigenratio} shows,
the lattice data agree well with the ChRMT predictions
(middle panel). 
It is also known that there exists the so-called
flavor-topology duality in ChRMT: the low-mode spectrum is
identical between the two-flavor (massless) theory at $Q=0$
and the quenched theory at $Q=2$ (right panel), while the
quenched spectrum at $Q=0$ is drastically different (left
panel). 
This is nicely reproduced by the lattice data.
These clear patterns indicate that the low-lying modes of
the QCD Dirac operator are in fact responsible for the
zero-momentum mode of the pion field $U$, which induce the vacuum with
spontaneously broken chiral symmetry.

If we look into the details, however, there is a slight
disagreement in higher eigenvalues. 
For instance, the deviation in the combination
$k/l$ = 4/1 is 10\%.
The reason for this will be discussed later.

\begin{figure}[tbph]
  \centering
  \includegraphics[width=8.5cm]{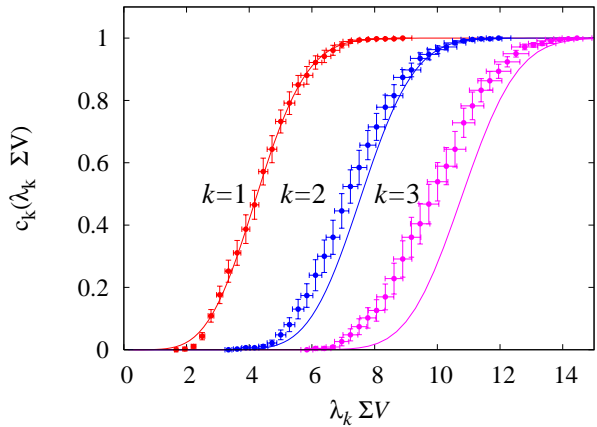}
  \caption{
    Cumulative distribution of the low-lying eigenvalues. 
    The horizontal error comes from the statistical error of
    $\Sigma$. 
    The solid curves are the ChRMT results with the input
    from the average $\langle\lambda_1\rangle$.
  }
  \label{fig:eigenhist}
\end{figure}
\begin{table}[tbp]
  \centering
  \begin{tabular}{ccccc}
    \hline\hline
    & $\langle\zeta_k \rangle$ 
    & $\langle\lambda_k\rangle \Sigma V$ 
    & $\langle\Delta\zeta_k \rangle$ 
    & $\langle\Delta\lambda_k\rangle\Sigma V$ \\
    \hline
    $k=1$ & 4.30 & [4.30] & 1.234 & 1.215(48) \\
    $k=2$ & 7.62 & 7.25(13) & 1.316 & 1.453(83) \\
    $k=3$ & 10.83 & 9.88(21) & 1.373 & 1.587(97) \\
    $k=4$ & 14.01 & 12.58(28) & 1.414 & 1.54(10) \\
    \hline
  \end{tabular}
  \caption{
    Comparison of the low-mode spectrum of ChRMT and the
    lattice data.
    The number given in $[]$ is used as an input.
  }
  \label{tab:low-modes}
\end{table}

Another non-trivial comparison can be made through
the shape of the eigenvalue distributions.
We plot the cumulative distribution 
$c_k(\zeta_k)\equiv\int_0^{\zeta_k}d\zeta^\prime p_k(\zeta^\prime)$ 
of the three lowest eigenvalues in Fig.~\ref{fig:eigenhist}.
The agreement between the lattice data and ChRMT
(solid curves) is quite good for the lowest eigenvalue.
For the higher modes the agreement of the central value
is marginal while the shape seems well described by ChRMT. 
Table~\ref{tab:low-modes} lists the numerical results of
both ChRMT and lattice data.
To characterize the shape of the distribution, 
we define the width 
$\langle\Delta\zeta_k\rangle \equiv
 \sqrt{\langle\zeta_k^2\rangle - \langle\zeta_k\rangle^2}$
as well as its lattice counterpart
$\langle\Delta\lambda_k\rangle\Sigma V$.
The overall agreement is very good, but we see 
deviations of about 10\% in the averages and 16\% in the
widths, at the largest. 

Because of the exact chiral symmetry in our simulation, we
do not expect significant effects due to finite $a$ in the
comparison of low-lying eigenvalues. 
The largest possible source of systematic errors is the
higher order effects in the $\epsilon$-expansion, which is a
finite volume effect.
Although the higher order corrections can not be calculated
within the framework of ChRMT, an estimate of their size can
be given by ChPT.
The next-to-leading order correction to the chiral
condensate is given by
$\Sigma[1+ ((N_f^2-1)/N_f) \beta_1/(FL)^2]$,
where $\beta_1$ is a numerical constant depending on the
lattice geometry \cite{Gasser:1987ah}.
Numerically, the correction is 13\% 
(we assume the pion decay constant $F$ = 93~MeV),
which is about the same size of the deviation of the
eigenvalue distributions.

Since the contribution from finite momentum modes is
  expected to be more significant for higher eigenvalues, we
  take the lowest eigenvalue as an input for the
  determination of $\Sigma$.
From the average of $\lambda_1$ we obtain
$\Sigma^{lat} a^3 = 0.00212(6)$ in the lattice unit and
$\Sigma^{lat}$ = [240(2)(6)~MeV]$^3$ in the physical unit. 
The second error in the latter comes from the lattice scale
$a$ = 0.107(3)~fm.
We put a superscript $\mathit{lat}$ in order to emphasize
that it is defined on the lattice.

In order to convert $\Sigma^{lat}$ to the definition in the
$\overline{\mathrm{MS}}$ scheme, we have calculated the
renormalization factor
$Z_S^{\overline{\mathrm{MS}}}(\mathrm{2~GeV})$
using the non-perturbative renormalization technique
through the RI/MOM scheme \cite{Martinelli:1994ty}.
Calculation is done at $m$ = 0.002 with several different
valence quark masses. 
The result is 
$Z_S^{\overline{\mathrm{MS}}}(\mathrm{2~GeV})$ = 1.14(2).
Details of this calculation will be presented elsewhere.

Including the renormalization factor, our final result is
$\Sigma^{\overline{\mathrm{MS}}}(\mathrm{2~GeV})$ = 
[251(7)(11)~MeV]$^3$.
The errors represent a combined statistical error 
(from $\lambda_1$, $r_0$ and $Z_S^{\overline{\mathrm{MS}}}$) and 
the systematic error estimated from the higher order effects
in the $\epsilon$-expansion, respectively.
Since the calculation is done at a single lattice spacing,
the discretization error cannot be quantified reliably, but
we do not expect much larger error because our lattice
action is free from $O(a)$ discretization effects.

In this letter we demonstrated that the lattice QCD
simulation is feasible near the chiral limit, as far as 
the exact chiral symmetry is respected.
The link between QCD and ChPT is established in the
$\epsilon$-regime without recourse to chiral
extrapolations. 
From their correspondence, the low energy constants, such
as the chiral condensate, can be precisely calculated.
Improvement of the precision may be achieved by increasing
the physical volume, as the higher order effects 
in the $\epsilon$-expansion are suppressed as $1/L^2$.
Further information on the low energy constants
can be extracted in the $\epsilon$-regime 
by calculating two- and three-point
functions or analyzing the Dirac eigenvalue spectrum with 
imaginary chemical potential
\cite{Damgaard:2001js, Hernandez:2006kz, Akemann:2006ru}.
This work is a first step towards such programs.

\begin{acknowledgments}
  We thank P.H.~Damgaard and S.M.~Nishigaki for fruitful
  discussions at the YITP workshop YITP-W-05-25 on
  ``Actions and Symmetries in Lattice Gauge Theory.''
  Numerical simulations are performed on IBM System Blue Gene
  Solution at High Energy Accelerator Research Organization
  (KEK) under a support of its Large Scale Simulation
  Program (No. 06-13).
  This work is supported in part by the Grant-in-Aid of the
  Japanese Ministry of Education 
  (No.~13135204, 1315213, 15540251, 16740156, 17740171, 18340075,
  18034011, 18740167, and 18840045)
  and the National Science Council of Taiwan (No. NSC95-2112-M002-005).
\end{acknowledgments}

\end{document}